\documentclass[12pt]{iopart}
\usepackage{graphicx}
\begin{document}

\title{Jet-hadron correlations in STAR}

\author{Alice Ohlson for the STAR Collaboration}

\address{Physics Department, Yale University, 272 Whitney Ave., New Haven, 06520-8120 CT, USA.}

\ead{alice.ohlson@yale.edu}

\begin{abstract}
Advancements in full jet reconstruction have made it possible to use jets as triggers in azimuthal angular correlations to study the modification of hard-scattered partons in the medium created in ultrarelativistic heavy-ion collisions.  This increases the range of parton energies accessible in these analyses and improves the signal-to-background ratio compared to dihadron correlations.  Results of a systematic study of jet-hadron correlations in central Au--Au collisions at $\sqrt{s_{NN}} = 200$ GeV are indicative of a broadening and softening of jets which interact with the medium. Furthermore, jet-hadron correlations suggest that the suppression of the associated hadron yield at high-$p_{T}$ is balanced in large part by low-$p_{T}$ enhancement.
\end{abstract}

\section{Introduction}
Jets can be used to study the energy loss of partons which traverse the hot and dense medium produced in ultrarelativistic heavy-ion collisions \cite{ref:whitepaper}.  Conceptually, models of partonic energy loss fall into two general categories: (1) radiative/collisional energy loss models \cite{ref:radcoll}, in which partons lose energy and are scattered as they traverse the medium (radiative energy loss is expected to dominate for light partons), and (2) ``black-and-white'' models, in which jets either exit the medium entirely unmodified or are completely thermalized.  In the former, azimuthal angular correlations would show a difference between the widths and yields of the jet peaks in Au--Au compared to p--p, while in the latter there would be no observed jet shape modification.  An analysis of jet shapes with jet-hadron correlations can distinguish between these classes of models.  Utilizing a reconstructed jet as a trigger increases the range of accessible parton kinematics compared to dihadron correlations \cite{ref:dihadron}.  

\section{Data sets and Analysis}
The data used in this analysis were collected by the STAR detector at the Relativistic Heavy Ion Collider (RHIC) in p--p and Au--Au collisions at $\sqrt{s_{NN}}$ = 200 GeV in 2006 and 2007.  Jets are reconstructed from charged tracks in the Time Projection Chamber (TPC) and neutral towers in the Barrel Electromagnetic Calorimeter (BEMC) using the anti-$k_{T}$ algorithm from the FastJet package \cite{ref:fj} with a resolution parameter $R = 0.4$.  Only tracks and towers with $p_{T}$\textgreater 2 GeV/$c$ are used in the jet reconstruction in order to minimize the effect of background fluctuations.  Furthermore, the reconstructed jets must include a BEMC tower that fired the online high tower (HT) trigger, which requires $E_T > 5.4$ GeV in one BEMC tower ($\Delta\varphi \times \Delta\eta = 0.05 \times 0.05$).  An offline software HT threshold of $E_T > 6$ GeV is imposed to reduce the effects of the trigger turn-on behavior.  

In order to analyze the jet signal in Au--Au collisions it is necessary to subtract the large heavy-ion background from the raw $\Delta\varphi = \varphi_{jet} - \varphi_{assoc}$ distributions (where the associated particles are all charged hadrons in the event).  Using the ZYAM (``Zero Yield at Minimum'') technique is unfavourable because it overestimates the background level in the presence of broad jet peaks.  Therefore in this analysis background levels are estimated by fitting the raw correlation function with the following functional form:  
\begin{equation}\label{eq:bkgfit}
\mbox{Nearside Gaussian} + \mbox{Awayside Gaussian} + B\left(1+2v_2^{assoc}v_2^{jet}\cos(2\Delta\varphi)\right)
\end{equation}
where $v_2^{assoc}$ is given by the standard STAR parameterization ($^1/_2(v_2\{2\}+v_2\{4\})$ as a function of $p_T$) \cite{ref:v2paper} and $v_2^{jet}$ is estimated to be $v_2\{2\}$ at $p_T = 6\mbox{ GeV}/c$.  Because $v_2^{jet}$ is not yet well constrained experimentally, the lower and upper bounds on the uncertainty of $v_2^{assoc}v_2^{jet}$ are conservatively estimated to be 0 and 150\% of $v_2\{2\}(p_T^{assoc}) \cdot v_2\{2\}(6\mbox{ GeV}/c)$.  The nearside and awayside associated hadron yields, denoted by $Y$, are given by the integrals of the nearside and awayside Gaussians (at $\Delta\varphi = 0$ and $\Delta\varphi = \pi$, respectively).  

The comparison between jets in Au--Au and p--p is quantified with four measures: the Gaussian widths of the jet peaks from (\ref{eq:bkgfit}), $I_{AA}(p_T^{assoc})$, $D_{AA}(p_T^{assoc})$, and $\Delta B$.  
\begin{equation}\label{eq:Iaa}
I_{AA}(p_{T}^{assoc}) = \frac{Y_{AA}(p_{T}^{assoc})}{Y_{pp}(p_{T}^{assoc})}
\end{equation}
\begin{equation}\label{eq:Daa}
D_{AA}(p_{T}^{assoc}) = Y_{AA}(p_{T}^{assoc}) \cdot \langle p_{T}^{assoc}\rangle_{AA} - Y_{pp}(p_{T}^{assoc}) \cdot \langle p_{T}^{assoc}\rangle _{pp}
\end{equation}
\begin{equation}\label{eq:deltaB}
\Delta B = \int d p_{T}^{assoc} D_{AA}(p_{T}^{assoc})
\end{equation}
$I_{AA}$, the ratio of the associated yields, folds in the shapes of the associated particle spectra in Au--Au and p--p, while $D_{AA}$ measures the energy difference between Au--Au and p--p.  If jets in Au--Au fragment like in p--p, then $I_{AA} = 1$ and $D_{AA} = 0$ for all $p_T^{assoc}$.  Deviations from these values are indicative of jet modification or unaccounted-for bulk effects (such as higher-order harmonics which have not been considered in this analysis).  Even if $D_{AA}(p_T^{assoc}) \neq 0$, it is possible for $\Delta B = 0$, indicating that the overall energy is balanced between high-$p_T^{assoc}$ and low-$p_T^{assoc}$ fragments.  

\section{Trigger Jet Matching}
It is expected that the HT trigger jet sample in Au--Au is highly biased towards unmodified jets, thus allowing easier comparison between jets in Au--Au and p--p.  However, on the nearside there are small deviations from $I_{AA} = 1$ and $D_{AA} = 0$.  The approach in this study is to assign maximum systematic uncertainties on the trigger jet energy scale corresponding to two extreme scenarios: the deviations from $I_{AA} = 1$ and $D_{AA} = 0$ are due entirely to (i) the bulk or (ii) jet modification.  
\begin{enumerate}
\item Under this assumption, a HT trigger jet in Au--Au is equivalent (at all $p_T^{assoc}$) to a p--p HT trigger jet embedded in a Au--Au minimum bias (MB) event, which provides a reasonable description of heavy-ion background fluctuations.  However, it is seen that the shape of the jet energy spectrum in Au--Au does not quite match the spectrum of p--p HT jets embedded in Au--Au MB events, even after accounting for detector effects.  The difference in spectra shapes is encompassed by a $\Delta E = -1$ GeV/$c$ uncertainty applied to the p--p trigger jet energy.  

\item Under this assumption, $\Delta B = 0$ when parton energies are correctly matched, even though $p_T^{jet,AA} \neq p_T^{jet,pp}$ because $p_T^{jet}$ is calculated only from charged tracks and neutral towers with $p_T > 2$ GeV/$c$.  The shift in the p--p trigger jet energy necessary to force $\Delta B = 0$ defines another systematic uncertainty ($+1.5 \cdot \Delta B$, where the factor of $1.5$ accounts for the neutral energy fraction).  
\end{enumerate}

These conservative estimates of the uncertainties on the trigger jet energy scale, which encompass all known possible sources of modification, are propagated to the awayside, where modification of the recoil jet due to jet-medium interactions can be studied.  

\section{The Recoil Jet}
\begin{figure}[ht]
\begin{minipage}{0.35\linewidth}
\includegraphics[width=80mm]{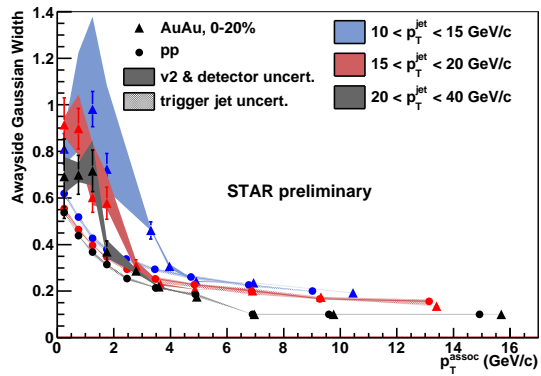}
\end{minipage}
\begin{minipage}{0.65\linewidth}\caption{\label{fig:asw}The Gaussian widths of the awayside jet peaks in Au--Au (triangles) and p--p (circles) indicate broadening of the awayside jet in Au--Au.}
\end{minipage}
\end{figure}

\begin{figure}[ht]
\begin{minipage}{0.5\linewidth}
\includegraphics[width=80mm]{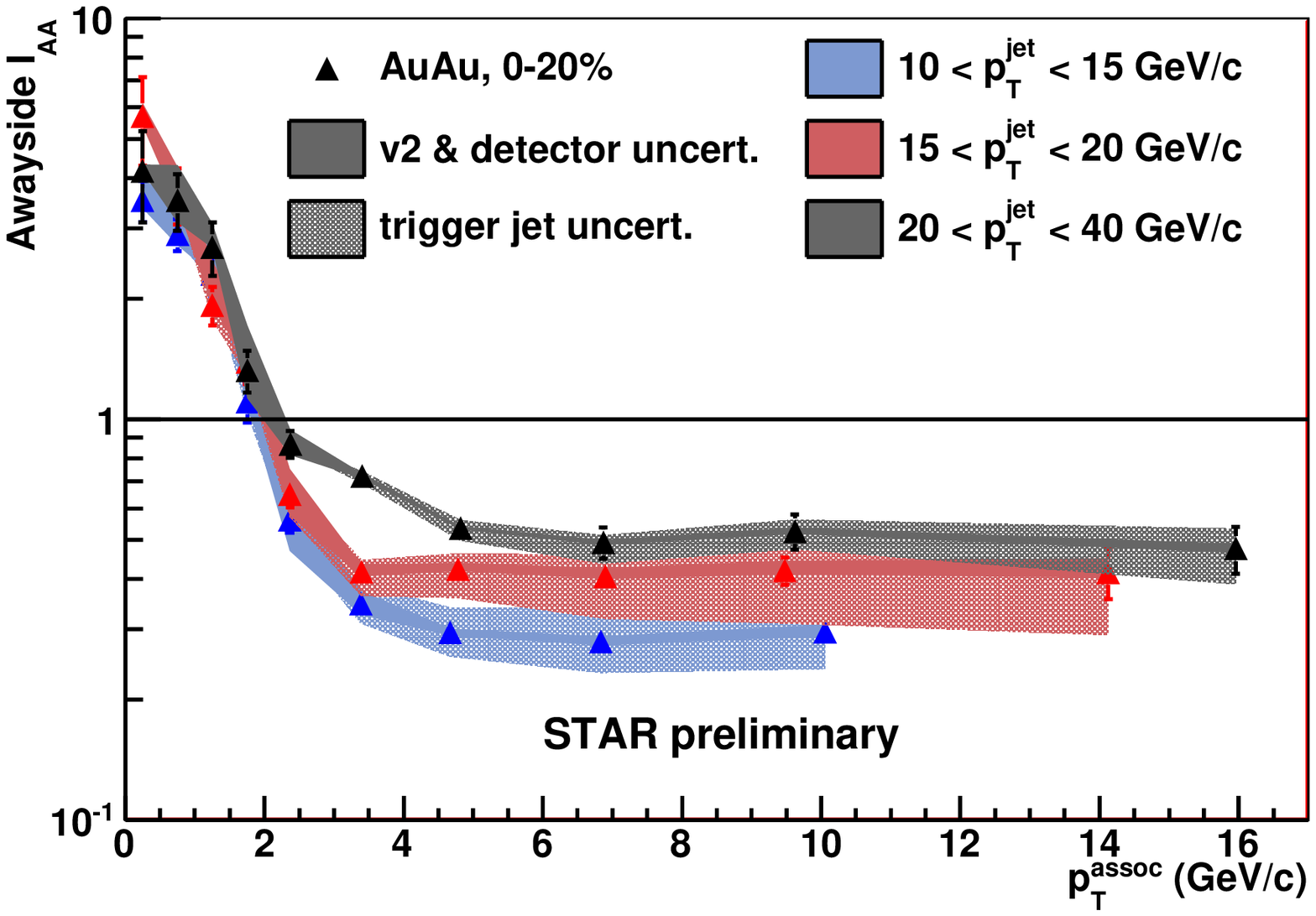}
\end{minipage}
\begin{minipage}{0.5\linewidth}
\includegraphics[width=80mm]{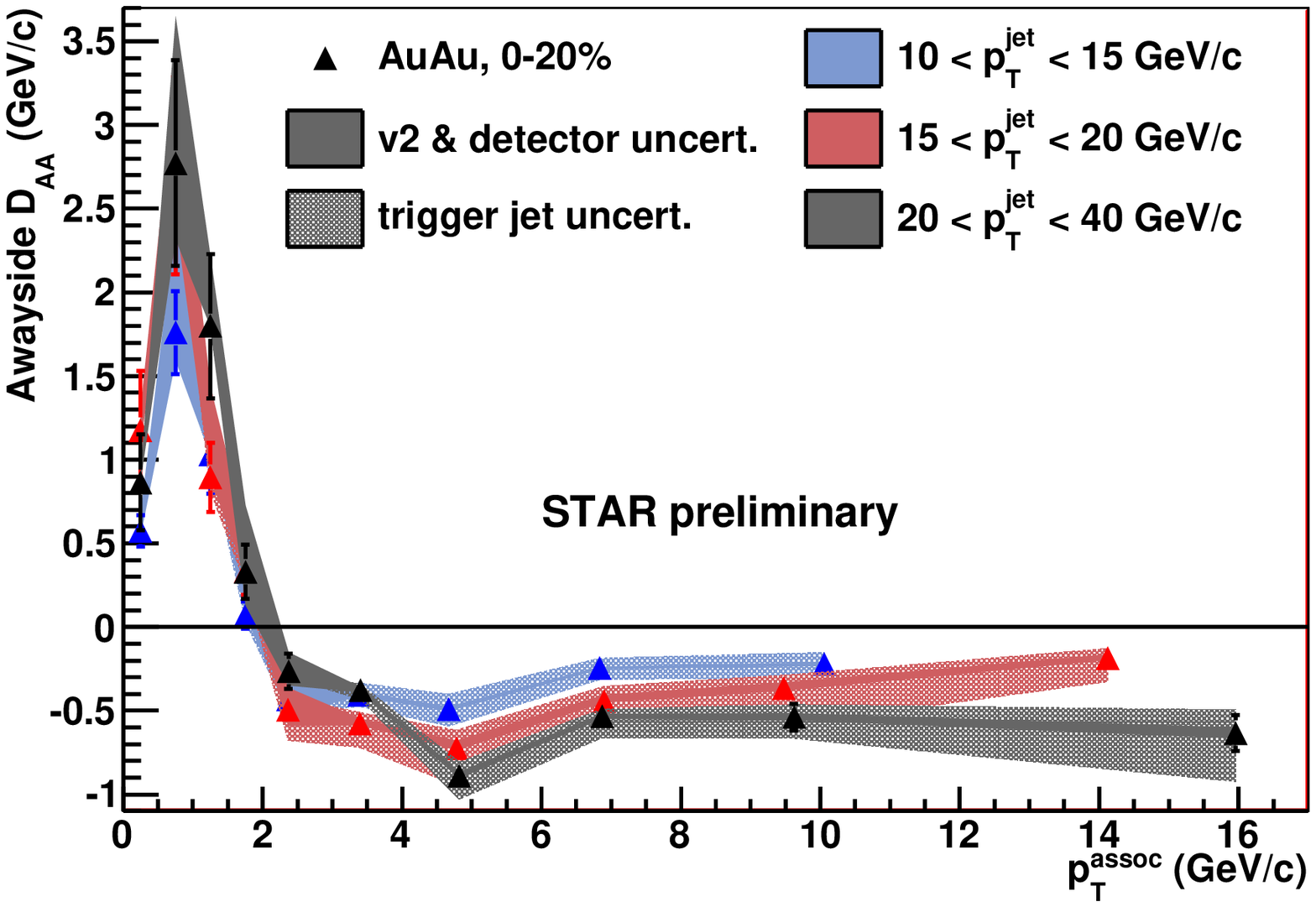}
\end{minipage}
\caption{\label{fig:asi}The awayside $I_{AA}$ (left) and $D_{AA}$ (right) indicate a softening of the awayside jet for three reconstructed jet energy ranges.  The awayside $D_{AA}$ shows that high-$p_T^{assoc}$ suppression is compensated for by low-$p_T^{assoc}$ enhancement.}
\end{figure}

The nearside jet is expected to have a surface bias \cite{ref:trenk} which will cause the recoil parton to travel through a significant amount of the medium, therefore maximizing any quenching or partonic energy loss effects on the awayside.  The awayside Gaussian widths (Figure \ref{fig:asw}) indicate significant broadening of the jet peak in Au--Au compared to p--p.  The awayside $I_{AA}$ and $D_{AA}$ (Figure \ref{fig:asi}) indicate significant softening of the awayside jets in Au--Au.  However, it is difficult to draw conclusions from the clear jet energy trends because the horizontal axes are $p_T^{assoc}$, not $z$, and because $I_{AA}$ depends on the shapes of the recoil spectra.  The values of $\Delta B$ in Table \ref{tab:DeltaB} show that a significant fraction of the high-$p_T^{assoc}$ suppression is being compensated for by low-$p_T^{assoc}$ enhancement.  All of these observations are indicative of significant jet quenching on the awayside.  

\begin{table}[ht]
\begin{minipage}{0.35\linewidth}
\begin{indented}
\item[]\begin{tabular}{@{}cc}
\br
$p_T^{jet}$ (GeV/$c$) & AS $\Delta B$ (GeV/$c$)\\
\mr
10-15 & $1.6^{+1.5+0.5}_{-0.3-0.5}$\\
15-20 & $2.3^{+1.8+0.5}_{-0.5-1.3}$\\
20-40 & $2.5^{+2.0+0.5}_{-0.8-0.8}$\\
\br
\end{tabular}
\end{indented}
\end{minipage}
\begin{minipage}{0.65\linewidth}
\caption{\label{tab:DeltaB}Awayside (AS) $\Delta B$ values for three $p_T^{jet}$ ranges.  The first set of systematic uncertainties is due to $v_2$ and detector uncertainties.  The second set of systematic uncertainties is defined by the $\Delta E$ and $\Delta B$ shifts.}
\end{minipage}
\end{table}

\section{Conclusions}
Jet-hadron correlations can be used to investigate the modification of partons that traverse the medium created in heavy ion collisions.  The results indicate that in high tower trigger events the recoil jet is significantly broadened and softened in Au--Au compared to p--p.  Furthermore, most of the high-$p_T$ suppression is balanced by low-$p_T$ enhancement.  The observed modifications in the jet shapes between Au--Au and p--p are in qualitative agreement with the radiative energy loss picture.  

\section*{Acknowledgments}
This  research  was  supported  in  part  by an  award from  the  Department  of  Energy  (DOE)  Office  of  Science  Graduate  Fellowship  Program administered  by  the  Oak  Ridge  Institute  for  Science  and  Education.  This research was also supported by the facilities and staff of the Yale University Faculty of Arts and Sciences High Performance Computing Center.

\bibliography{references}{}

\section*{References}
\begin{thebibliography}{9}
\bibitem{ref:whitepaper} Adams J et al. 2005 \textit{Nucl. Phys.} A \textbf{757} 102
\bibitem{ref:radcoll} Majumder A and van Leeuwen M 2011 \textit{Prog. Part. Nucl. Phys.} \textbf{66} 41
\bibitem{ref:dihadron} Aggarwal MM et al. 2010 \textit{Phys. Rev.} C \textbf{82} 024912
\bibitem{ref:fj} Cacciari M and Salam G 2006 \textit{Phys. Lett.} B \textbf{641} 57
\bibitem{ref:v2paper} Adams J et al. 2005 \textit{Phys. Rev.} C \textbf{72} 14904
\bibitem{ref:trenk} Renk T and Eskola K 2007 \textit{Phys. Rev.} C \textbf{75} 054910
\end{thebibliography}
\end{document}